\documentclass{article}

\usepackage{PRIMEarxiv}

\usepackage[utf8]{inputenc} 
\usepackage[T1]{fontenc}    
\usepackage{hyperref}       
\usepackage{url}            
\usepackage{booktabs}       
\usepackage{amsfonts}       
\usepackage{nicefrac}       
\usepackage{microtype}      
\usepackage{lipsum}
\usepackage{fancyhdr}       
\usepackage{graphicx}       
\graphicspath{{media/}}     

\title{Investigating Software Testability and Test cases Effectiveness}

\author{
  Mamdouh Alenezi \\
  College of Computer and Information Sciences \\
  Prince Sultan University \\
  Riyadh, Saudi Arabia\\
  \texttt{malenezi@psu.edu.sa}
}

\begin{document}
\maketitle

\begin{abstract}
Software measurement is an essential management tool to develop robust and maintainable software systems. Software metrics can be used to control the inherent complexities in software design. To guarantee that the components of the software are inevitably testable, the testability attribute is used, which is a sub-characteristics of the software's maintabilility as well as quality assurance. This study investigates the relationship between static code and test metrics and testability and test cases effectiveness. The study answers three formulated research questions. The results of the analysis showed that size and complexity metrics are suitable for predicting the testability of object-oriented classes.
\end{abstract}

\keywords{Software Engineering \and Software Quality \and Testability}

\section{Introduction}
Software quality is a multi-dimensional concept in software engineering. Software testability is one characteristic of software quality that reflects the ease of testing a software artifact. Testability increases the chance of easily find faults in software artifacts such as modules, requirements, or design documents \cite{kasisopha2020method}. ISO/IEC 9126-1 standard has regarded testability as the ability of software to facilitate the modified software for validation \cite{iso2001iec}. ISO/IEC 25010:2011 Systems and software engineering — Systems and software Quality Requirements and Evaluation (SQuaRE) — System and software quality models \cite{garousi2019survey} discuss testability as one of the sub-characteristics of maintainability.

We can look at testability as the needed effort to test a module or a component of the software to ensure that the wanted function is accurately delivered \cite{filho2020correlations}. Testability is regarded as the comfort through which testing of the software can be performed and also recognizes some of the quality attributes such as observability and operability, which may direct towards the testability extent of the software. Testing is an essential activity in developing software systems to evaluate their quality \cite{filho2020correlations}. However, testing is a very expensive costly activity that requires special skills and tools \cite{winters2020software}. Testing involves ensuring several quality attributes are met, including performance, security, usability, etc. Developing a testable software system will surely reduce the cost of testing.

In case the testability is high of a specific part of the software system will surely make finding faults in the system easier. Having a lower level of testability will increase the testing effort, which will automatically decrease the chances of finding software defects \cite{voas1995software}. Every aspect that makes testing harder/slower will result in increasing the chances of having more bugs. Increasing the level of testability in the software will reduce the cost, increase quality, and produce higher-quality software \cite{garousi2019survey}.

Numerous studies presented different attributes that can influence testability \cite{freedman1991testability,binder1994design,gao2005component,filho2020correlations}.  According to \cite{filho2020correlations}, five main factors can affect testability, namely: comprehensibility, observability, controllability, traceability, and test support capability. McCall’s quality model considers testability as a quality factor of the product revision. This quality factor has quality criteria which are simplicity, instrumentation, self-descriptiveness, and modularity. Boehm’s quality model considers testability as an intermediate-level characteristic that is under the high-level characteristics (maintainability). The primitive characteristics listed under testability are accountability, communicativeness, self descriptiveness, and structuredness \cite{al2010quality}.

In this work, the author investigates software testability and test cases effectiveness. Three research questions have been formulated and studied. The primary significance of this study is investigating the role static code and test metrics can be played in predicting test-case effectiveness and testability—from mutation analysis perspectives.

\section{Software Testability}

Testability is one of the software quality models' characteristics and it is a very challenging endeavor to measure it and evaluate it \cite{mouchawrab2005measurement}. The challenge comes from the many potential factors that might affect testability. However, there are several reasons to measure and evaluate testability. Low testability software products are less trustworthy, even after successful testing \cite{485220}. Low testability components are more expensive to maintain and update especially when discovered late in the development process. On the other hand, good testability components can significantly increase software quality and reduce testing cost \cite{10.5555/861448}. Several research studies related testability and test efficiency to the testing cost and effort \cite{mouchawrab2005measurement,10.5555/861448,sej.1990.0011}.

Improving testability has a significant impact on making the testing more efficient \cite{485220} since testing can cost up to 50\% of the total software development cost and effort \cite{10.5555/2161638}. The recommended time spent by developers in writing tests should be between 25\% and 50\%. Further, modern software development methodologies emphasize the importance of testing such as agile, eXtreme Programming (XP) and Scrum \cite{4145039}. Test-Driven Development (TDD), as an example, ensures that tests are designed and developed before writing the code which assures that these components work correctly. In these processes, unit tests written by the developers/tester are integral parts of the software \cite{10.1007/3-540-47993-7_10}. These unit tests validate existing features while developing new functionality \cite{10.1007/3-540-47993-7_10}. Well-designed unit tests are a great aid in improving software quality and detecting bugs early in the development process \cite{1251026}. Usually, the right ratio of test code to production code is estimated to be 1:1 \cite{Deursen01refactoringtest}.

In summary, improving software testability has a positive direct impact on the overall software quality. However, measuring, evaluating, and assessing software testability remains a challenge.

\subsection{Defining Testability}

Significant research efforts were dedicated to defining and measuring testability. Several standards and research studies have different ways of defining testability. The IEEE standard glossary of software engineering terminology \cite{159342} has two definitions for testability, (1) “the degree to which a system or component facilitates the establishment of test criteria and the performance of tests to determine whether those criteria have been met”, and (2) “the degree to which a requirement is stated in terms that permit establishment of test criteria and performance of tests to determine whether those criteria have been met”. The first definition focuses on test criteria, whereas the second focuses on test coverage. The ISO standard \cite{ISO9126} describes testability as “attributes of software that bear on the effort needed to validate the software product”. Consequently, ISO sees testability from the effort needed to test a software product, while IEEE sees testability from a test criteria point of view.

McCall quality model \cite{al2010quality} was the first quality model as an attempt to emphasize software quality. McCall decomposed Testability to Simplicity, Instrumentation, and Modularity and defined it as the ability to validate requirements. Boehm \cite{10.5555/800253.807736} improved the McCall model by adding many factors at different levels. Boehm defined Testability to be how easy to validate that requirements meet the implemented software. The ISO 9126 quality model \cite{ISO9126} defined Testability as the effort needed for validating the software product.

Bache and Mullerburg \cite{bache1990measures} defined testability as “the minimum number of test cases to provide total test coverage, assuming that such coverage is possible”. The authors built their measurement on a static flow graph model. Freedman \cite{freedman1991testability} goal was to produce easily testable software by defining testability based on effort, time, and resources required for software testing. This discussion demonstrates that there are different views of software testability and its measurement and evaluation.

\subsection{Software Metrics and Measurement}

In any engineering discipline, measurement is an essential process. One of the main goals of software measurement is to help software engineers and project managers make informed predictions and decisions to support planning, monitoring projects to enable control, and judge the software project performance \cite{46876}. Software metrics can be used to quantify and measure several software characteristics. Software metrics are an essential part of software quality \cite{e9292cadbbd244e1993beacd21c53aa1} and have played an important role in analyzing and improving software quality \cite{544352}. In addition, software metrics can help ensure that quality requirements are actually being met.

Software metrics can be classified into three categories: product, process, and project metrics \cite{10.5555/1875255}. Quality has been usually in the literature associated with process and product metrics. Within these lines, Kan \cite{10.5555/559784} categorized quality measurements into two categories: in-process and end-product metrics. In-process metrics characterize the development process quality, while end-product metrics reveal the software product characteristics. Product metrics can be further categorized into static and dynamic metrics based on how they gauge different aspects of a system. The ISO 9126 model \cite{iso2001iec} reveals more details about these two types of software product metrics and how they measure software product quality:

\begin{itemize}
    \item Static metrics can measure attributes and properties of the software without the need to execute or run the program. They usually characterize the internal quality of the software.
    \item Dynamic metrics can measure attributes and properties of the software only at run-time.
\end{itemize}

As we are focusing on this research on static metrics only, we will explain static metrics even further. Static metrics measure non-running system representations which capture the static structure of a system  \cite{10.5555/2851535}. Static analysis techniques are used to collect such metrics to enable the exploration and analysis of the source code \cite{4145039}. Most existing quality metrics come in a static form such as the Lines of Code (LOC) metric, and McCabe’s Cyclomatic Complexity (CC) metric.

\section{Related Work}

This section is aimed at analyzing several research works that discussed testability measurement and analysis. The work \cite{garousi2019survey} executed a systematic review of the past literature regarding software testability between 1982 and 2017. They highlighted several classifications and aspects of the research done during that period.

Binder \cite{binder1994design} discussed the importance of testability and how it affects revealing software faults. It was criticized by Binder that six chief elements impacted the testability, which include the requirement characteristics, implementation characteristics, testing ability, environment where the test is undertaken, suite of the test, as well as the entire development process of the software. The research pinpointed multiple metrics of the source code for analyzing the testability. The final outcome indicated some extent of a strong association of these metrics found with testability. Bruntink and van Deursen \cite{bruntink2004predicting} projected testability as the total number of test cases as well as the efforts and dedication put in to generate those test cases. They also investigated the association between the testing effort required for the unit tests and the metrics of the static code. The findings revealed that the module size has a significant impact on the effort of the testing. Badri and Toure \cite{badri2012empirical} made use of the logistic regression method for the development of the association between the testing effort required for the classes and the metrics of the software. Their findings highlighted that the coupling, complexity, metrics size, and cohesion could be utilized for estimating the class's efforts to testing. 

Voas and Miller \cite{voas1995software} performed research to calculate the testability making use of the sensitivity analysis that calculates for a specific location within a program the likelihood of the failure, which could be incorporated into the program via a single fault. Bache and Mullerburg \cite{bache1990measures} also analyzed the testability constituting the minimum number of test cases needed for complete testing coverage. Their study was grounded on the coverage of the control flow.

Baudry et al. \cite{baudry2002testability} proposed to measure testability using object-oriented classes interactions in UML class diagrams. In the case of having two or more distinct paths between two classes means more testing effort is needed. Mouchawrab et al. \cite{mouchawrab2005measurement} used a hierarchical OO software testability framework to measure testability from UML diagrams. UML relationships can be used to assist in calculating or measuring the testability at the early stages of the development process of the software. They demonstrated that different object-oriented coupling metrics are controlled by the power-law distributions. Jungmayr \cite{jungmayr2002identifying} looked at testability as the level of dependencies between classes. He defined testability as having more dependencies means more test cases are needed. He recommends the perception of the test-critical metrics and dependencies to recognize them via refactoring. 

Briand et al. \cite{briand2003investigating} proposed the use of class invariants (contracts) to improve testability. Contracts can improve testability by increasing the likelihood of detecting and identifying the location of faults. Khatri et al. \cite{khatri2013improving} proposed a new methodology to improve testability using software reliability growth models. Though several studies have examined the relationship between software metrics and testability, there is no conclusive position on the applicability of design metrics in measuring the testability of software. Thus, more studies are required to further investigate the suitability of metrics as testability indicators. In our work, we focus on static code and test metrics since they can be easily computed with the need to running the software.

\section{Research Methodology}

The key findings mentioned in this section focuses on the dataset used in the research, the three formulated research questions, and the methods used to answer these questions.

\subsection{Dataset}

This work uses one of the available testability datasets \cite{terragni2020measuring}. The data is from 1,186 Java open-source projects of different sizes and categories. The data contains 28 class metrics across different categories (inheritance, size, complexity, cohesion, encapsulation, and coupling), 6 test effort metrics, and 3 test quality metrics. Precise information of the collected data and tools for analyzing and calculating the metrics are mentioned in \cite{terragni2020measuring}. Table \ref{table:1} shows the details of the data.

\begin{table}[]
\centering
\begin{tabular}{|l|l|l|} \hline
Metric Type          & Design Property & Metric                                    \\ \hline
Code Metrics         & Size            & Lines of Code (LOC)                       \\
                     &                 & Number of Bytecode Instructions (NBI)     \\
                     &                 & Lines of Comment (LOCCOM)                 \\
                     &                 & Number of Public Methods (NPM)            \\
                     &                 & Number of Static Method (NSTAM)           \\
                     &                 & Number of Fields (NOF)                    \\
                     &                 & Number of Static Fields (NSTAF)           \\
                     &                 & Number of Method Calls (NMC)              \\
                     &                 & Number of Method Calls Internal (NMCI)    \\
                     &                 & Number of Method Calls External (NMCE)    \\ \cline{2-3}
                     & Complexity      & Weighted Methods per Class (WMC)          \\
                     &                 & Average Method Complexity (AMC)           \\
                     &                 & Response for a Class (RFC)                \\\cline{2-3}
                     & Inheritance     & Depth of Inheritance Tree (DIT)           \\
                     &                 & Number of Children (NOC)                  \\
                     &                 & Measure of Functional Abstraction (MFA)   \\\cline{2-3}
                     & Coupling        & Coupling Between Object classes (CBO)     \\
                     &                 & Inheritance Coupling (IC)                 \\
                     &                 & Coupling Between Methods (CBM)            \\
                     &                 & Afferent Coupling (Ca)                    \\
                     &                 & Efferent Coupling (Ce)                    \\\cline{2-3}
                     & Cohesion        & Lack of Cohesion in Methods (LCOM)        \\
                     &                 & Lack of Cohesion of Methods (LCOM3)       \\
                     &                 & Cohesion Among Methods in class (CAM)     \\\cline{2-3}
                     & Encapsulation   & Data Access Metrics (DAM)                 \\
                     &                 & Number of Private Fields (NPRIF)          \\
                     &                 & Number of Private Methods (NPRIM)         \\
                     &                 & Number of Protected Methods (NPROM)       \\\hline
Test Effort Metrics  &                 & Test – Lines of Code (T-LOC)              \\
                     &                 & Test – Number of Tests (T-NOT)            \\
                     &                 & Test – Number of Assertions (T-NOA)       \\
                     &                 & Test – Number of Method Calls (T-NMC)     \\
                     &                 & Test – Weighted Methods per Class (T-WMC) \\
                     &                 & Test – Average Method Complexity (T-AMC)  \\\hline
Test Quality Metrics &                 & Line coverage (L)                         \\
                     &                 & Branch coverage (B)                       \\
                     &                 & Mutation score (M)                       \\ \hline
\end{tabular}
\caption{\fontsize{10pt}{11pt}\selectfont{\itshape{Available metrics in the used Dataset}}}
    \label{table:1}
\end{table}

\subsection{Dataset Preprocessing}

The data has 9861 instances (classes) from 1,186 Java open-source projects. To formulate the data for these experiments, we did some preprocessing steps to make the data ready. We removed the first five columns since they have only information about the project name, URL, commit, and class and tests paths. Since the focus of our work is to investigate test effectiveness, we base our measurement on Mutation Score (M). The mutation score is computed as the killed mutants by the suite test percentage. In addition to evaluating the capability of the tests by the mutation scores to perform the seeded faults, the scores are also utilized to signify the capability of the oracles of the tests to expose those discrepancies. Mutation score values range from 0-1.

We classify tests in the dataset according to their mutation score as effective and not effective, making use of the mutation score quartiles. Tests that classify within the 1st quartile are allocated to the non-quartile set while the others within the 4th quartile into the effective set. Any test cases falling between these two quartiles were discarded. These estimations were in line with the previous research studies in the field of software engineering \cite{shirabad2000supporting,tian2015characteristics} in response to the discretization noise (data points bot allocated to specific classes) \cite{sun2017revisiting}. As the emphasis is laid on the exploration of the effective as well as non-effective test characteristics, so we remove test classes having average values of effectiveness.

After calculating the quartiles for this dataset, we found that test cases that have a mutation score of 1 are considered part of the fourth quartile, and 0.4 and less are considered part of the first quartile. We removed all instances that are in the remaining quartiles since they are considered average. The remaining instances are in the dataset are 5022. The study aims to evaluate the test effectiveness (testability) of Java object-oriented classes. To be able to run different experiments and answer the targeted research questions, dependent and independent variables should be specified.

\subsection{Dependent and Independent Variables}

Mutation score is used as the dependent variable, which is the percentage of the mutants killed divided by the number of mutants generated \cite{coles2016pit}. The selection is adherent to the past research within the domain of software testing, which reports the score of mutation to be the predominant criterion for the coverage of the code \cite{jia2010analysis} and one of the key indicators used by the developers \cite{andrews2005mutation,just2014mutants}. 

In the context of this study with regards to independent variables, an aggregate of 34 factors in addition to 2 chief dimensions (test-effort and code metrics) are considered. These metrics provide us with validation of whether the effectiveness of the test can be associated with these elements. One of the study objectives is to determine the lightweight estimation model depending only on the static code-quality characteristics, which may easily be computed onto the present version of the test classes. 

The set of code metrics constituted of 28 features estimating the inheritance, complexity, coupling, size, encapsulation, and cohesion. Such metrics are commonly used in previous studies. The notion for utilizing these metrics is dual: more and larger complicated production classes could be difficult to test and resulting, writing effective test cases for these classes could be difficult, and complicated and large test cases could cause deeper exhibit the code under test directing towards a high fault detection ability. 	

The calculation of the testing class effort is known as test-effort metrics with respect to the complexity and size of the aligned test cases. The effort of the test can be adequately utilized for the complexity and size of the test-effort metrics (test class) \cite{terragni2020measuring}. Metrics that we have used in the present study include the test numbers., line code of test, assertation number, method call number, methods of test weighted per class, cyclomatic complexity sum test, as well as the test average method complexity. 

\subsection{Research Questions}

The primary goal of this research is to develop an understanding regarding the factors which influence the test cases' effectiveness (mutation scores), such as the capability to detect faults, to formulate an automated strategy supporting the developers when analyzing how effective their developed test cases are. To attain this goal, we devised three research questions (RQs). Foremost, RQ signified a preliminary evaluation of the association existing within the 34 independent variables chosen and the effectiveness of the test cases. The aim is to develop an understanding of whether and to what degree the independent variable distribution of the values varies for test cases with high or low scores of the mutation. The second RQ facilitates in coming up with automated tactics, on the basis of Machine Learning (ML) strategies, to examine the effectiveness of a test on the basis of the score of the mutation. The last RQ further investigates these factors as we want to develop an understanding of the most significant features encapsulated by the proposed approach. This would aid in offering extended information for the developers concerned with the source code aspects to maintain control while forming a test case effectively. 
\begin{itemize}
    \item \textbf{RQ1}: Is there a relationship between the static metrics and test-case effectiveness?
    \item \textbf{RQ2}: To what extent do static code and test metrics helps in predicting the effectiveness of test cases?
    \item \textbf{RQ3}: What are the most influential testability metrics?
\end{itemize}

\subsection{Approach}

To answer RQ1, Spearman’s Rank Correlation Coefficient was conducted in this study. The rationale for choosing is to assess the statistical dependence of the two metrics. The benefit of using this tool is that it does not require a normal data distribution.

To answer RQ2, Different ML algorithms exist, yet not one of them is suitable for all the tasks. As the classifier made use of the prediction purposes that could highly impact the performance of the model, multiple classifiers were tested prior to choosing the best fit for the proposed estimation model. Hence, a series of ML algorithms were chosen from two contrasting yet well-established families, namely functional-based and tree-based techniques \cite{nasrabadi2021learning}, to determine the testability prediction model.
\begin{itemize}
    \item \textbf{Decision Tree} makes use of the highly informative characteristic to divide the instances with varying classes from one another. The practical relevance of the C4.5 decision tree version is widely accepted as it makes use of the concept underlying information entropy to choose the root nodes for every sub-tree. Decision trees are easy to comprehend and interpretable classifiers.
    \item \textbf{Random Forest} is a collaborative method of learning, creating a multivariate of the decision trees over the bootstrapped data. Every tree is constructed over a sub-set of the authentic data and exhibits feature outcomes in a highly robust classifier obtaining the most votes by a simple majority is a predicted class. Out-of-the-box performance of random forest is well reported.
    \item \textbf{Multilayer Perceptron} is a category of feed-forward neural networks embedded in a deep architecture that comprises different layers of computations units interlinked in a directed acyclic graph (DAG). This provides a feature of mapping input-to-output and works better in regression projects. MLP characteristics are inserted into the foremost network layer, and the outcome is developed in output, whereas the actual output is projected to be developed.
\end{itemize}
 
Features, in our case, the static metrics can be ranked according to their contribution to distinguish different class labels. Feature ranking algorithms assess individual features by assigning a score and rank them according to that score. To answer RQ3, we use four different algorithms to know the most influential metrics to testability namely, Gain Ratio, Information Gain, Symmetric Uncertainty, and OneR \cite{malhotra2021threshold}.

\section{Experimental Evaluation}

For evaluation purposes, Spearman’s Rank Correlation Coefficient was conducted in this study. We only report correlation above 0.5 positively or negatively. Table \ref{table:2} shows the correlation values. The table clearly shows that size and coupling are highly negatively correlated with the mutation score. To answer RQ1, there is a clear negative relationship between effective test cases and the complexity and size metrics of the class under test. This helps practitioners to decrease the class size and reduce the complexity of the class, which will greatly help in making the class more testable. 

\begin{table}[ht]
    \centering
    \setlength\tabcolsep{6pt}
    \begin{tabular}{ |l|c| }
        \hline
        \bfseries Static Metric & \bfseries Correlation Coefficient  \\ 
        \hline
        Number of Bytecode Instructions (NBI) &	-0.61652431  \\ 
        \hline
        Response for a Class (RFC)&	-0.603928734 \\ \hline
        Lines of Code (LOC)&	-0.58334397 \\ \hline
        Weighted Methods per Class (WMC)&	-0.56027992 \\ \hline
        Number of Method Calls (NMC)&	-0.55492337 \\ \hline
        Number of Method Calls External (NMCE)&	-0.512557727 \\ \hline
        Lines of Comment (LOCCOM)&	-0.50032704 \\ \hline
        Cohesion Among Methods in class (CAM)&	0.521969871 \\ \hline
    \end{tabular}
    \caption{\fontsize{10pt}{11pt}\selectfont{\itshape{The Spearman correlation of static metrics with the mutation score}}}
    \label{table:2}
\end{table}

To answer RQ2, we want to know to what extent static code and test metrics help in predicting the effectiveness of test cases. Table \ref{table:3} illustrates the chosen classifier's performance for the five selected evaluation criteria. As depicted, the model making use of the test metrics and static code exhibit high performance, not only relying on the F-Measure (87\%) but also during the AUC analysis. From the three selected classifiers, Random forest has taken the lead on all five evaluation criteria. These outcomes predominantly recommend that ML methods could be efficiently adhered to to evaluate the effectiveness of the test case. It is worth mentioning that we opted for projects from multiple domains; hence, we are assertive that our estimation model may be used generally in multiple contexts. In a nutshell, estimation and prediction models are efficient for the categorization of the effectiveness of the test cases. A model depending on static information accomplishes performance nearing 94\% with respect to AUC, making it a high-performing and more relevant solution for a real-life scenario. 

\begin{table}[ht]
    \centering
    \setlength\tabcolsep{6pt}
    \begin{tabular}{ |l|c|c|c|c|c|}
        \hline
        \bfseries Classifier & \bfseries Accuracy & \bfseries Precision &	\bfseries Recall & \bfseries F-Measure & \bfseries AUC \\ \hline
        Random Forrest & 0.877 & 0.879 & 0.877 & 0.877 & 0.942 \\ \hline
        Multilayer Perceptron&	0.830&	0.830&	0.830&	0.830&	0.895 \\ \hline
        Decision Tree&	0.825&	0.827&	0.826&	0.826&	0.870 \\ \hline
    \end{tabular}
    \caption{\fontsize{10pt}{11pt}\selectfont{\itshape{The classification results of the prediction model}}}
    \label{table:3}
\end{table}

To answer RQ3, we compare different ranking algorithms to know the most testability influential metrics. Table \ref{table:4} shows the most important static metrics in terms of predicting the effectiveness of the studied test cases. These can be considered good indicators of measuring the testability of classes. We can see from Table 4 that the first ranked metrics are found by all selected algorithms. These important metrics are NBI, RFC, LOC, LOCCOM, WMC, and NMC. We can further explore these selected metrics by classifying them to design properties. Number of Bytecode Instructions (NBI), Lines of Code (LOC), Lines of Comment (LOCCOM), and Number of Method Calls (NMC) are considered code size metrics. Whereas Response for a Class (RFC) and Weighted Methods per Class (WMC) are considered complexity metrics. This clearly indicated that these metrics need to be watched and monitored when developing software systems to ensure system testability.

\begin{table}[ht]
    \centering
    \setlength\tabcolsep{6pt}
    \begin{tabular}{ |l|c|c|c|c|}
        \hline
        \bfseries   & \bfseries Gain Ratio & \bfseries Information Gain &	\bfseries Symmetrical Uncertainty & \bfseries OneR \\ \hline
        Rank 1&	NBI&	NBI&	NBI&	RFC \\ \hline
        Rank 2&	RFC&	RFC&	RFC&	NBI \\ \hline
        Rank 3&	LOC&	LOC&	LOC&	LOC \\ \hline
        Rank 4&	LOCCOM&	NMC&	LOCCOM&	NMC \\ \hline
        Rank 5&	WMC&	WMC&	NMC&	LOCCOM \\ \hline
        Rank 6&	NMC&	LOCCOM&	WMC&	WMC \\ \hline
        Rank 7&	NMCE&	CAM&	NMCE&	CAM \\ \hline
        Rank 8&	CAM&	NMCE&	CAM&	NMCE \\ \hline
        Rank 9&	T-NMC&	NPM&	NPM&	NPM \\ \hline
        Rank 10& NPM&	LCOM&	LCOM&	LCOM \\ \hline
    \end{tabular}
    \caption{\fontsize{10pt}{11pt}\selectfont{\itshape{Important metrics using different feature ranking algorithms}}}
    \label{table:4}
\end{table}

\section{Discussion}

One of the key findings of this research is the role of source code quality to address the test cases' effectiveness. Preventive information can be provided to the software practitioners regarding the presence of the non-effective test. As the model comprises quality-based characteristics, non-effective tests can be emphasized by the developers to enhance their design. Additional devotion by the practitioner towards the maintenance effort to develop the test more focused on the production code to enhance its design and improve its testability. 

The recommended estimation model can complement the process of checking the quality of the testing process. The identification of a non-effective test might be of interest for the developer for assessing the test issues by running already established toold for mutation testing which offer a comprehensive insight of the causes stopping the test from catching any pertinent faults. The initial estimation of the effectiveness of the test code can facilitate in comprehending the factors related to quality which influence the effectiveness of the test. 

\section{Validity Threats}

This section addresses few factors influencing the validity of our experiments and result.  

\textbf{Threats to construct validity}. The chief threat is the possible imprecision in the data extraction and analysis process in the used dataset. The data were collected according to the best practices and widely used tools. We estimated the effectiveness of the test case, making use of the mutation score, depending on the supposition that mutation score is a good representative of quality of test cases. To analyze the abilities and characteristics of the estimated models for the prediction of the non-effective and effective tests, tests with average effectiveness (also known as the discretization noise) were excluded. 

\textbf{Threats to conclusion validity}. To explore the difference existing within the factors of the non-effective and effective tests. When analyzing the effectiveness of the test case, we compared multiple classifiers. To choose and authenticate the most appropriate model, we utilized a cross-validation process with 10-folds. Also, we utilized different evaluation metrics to offer a diverse insight regarding the devised model performance. Lastly, when examining the highly relevant features, we utilize different feature ranking along with selection algorithms. 

\textbf{Threats to external validity}. Twenty-eight factors were considered divided into two distinct categories. The effectiveness of the test case may be influence by some additional factors which were not part of this study. We plan to improve the set of factors in our project's future works. In terms of the experiment size, we explored a dataset comprising of 1,186 Java open-source projects of varying categories and sizes. Though it already exhibits a large-scale empirical experiment, there are still further replications to target multiple project types.

\section{Conclusions and Future Work}

In this work, the author studied the relation between 28 factors (test metrics and static code) and the effectiveness and testability of the test cases. Then, a prediction model is developed to use the already established factors to differentiate the non-effective and effective tests. Moreover, this research emphasizes the most relevant and essential code metrics, which highlight the testability of the object-oriented classes. In the future, further experiments will be conducted to explore how additional metrics impact the test case effectiveness, support multiple programming languages, and how these insights can be utilized for other testing-related tasks like the selection of the test case, prioritization, and minimization.

\bibliographystyle{unsrt}  
\bibliography{references}

\begin{thebibliography}{10}

\bibitem{kasisopha2020method}
Natsuda Kasisopha, Songsakdi Rongviriyapanish, and Panita Meananeatra.
\newblock Method evaluation for software testability on object oriented code.
\newblock In {\em 2020 59th Annual Conference of the Society of Instrument and
  Control Engineers of Japan (SICE)}, pages 308--313. IEEE, 2020.

\bibitem{iso2001iec}
ISO Iso.
\newblock Iec 9126-1: Software engineering-product quality-part 1: Quality
  model.
\newblock {\em Geneva, Switzerland: International Organization for
  Standardization}, 21, 2001.

\bibitem{garousi2019survey}
Vahid Garousi, Michael Felderer, and Feyza~Nur K{\i}l{\i}{\c{c}}aslan.
\newblock A survey on software testability.
\newblock {\em Information and Software Technology}, 108:35--64, 2019.

\bibitem{filho2020correlations}
Francisco Gutenberg~S Filho, Val{\'e}ria Lelli, Ismayle de~Sousa Santos, and
  Rossana~MC Andrade.
\newblock Correlations among software testability metrics.
\newblock In {\em 19th Brazilian Symposium on Software Quality}, pages 1--10,
  2020.

\bibitem{winters2020software}
Titus Winters, Tom Manshreck, and Hyrum Wright.
\newblock {\em Software engineering at google: Lessons learned from programming
  over time}.
\newblock O'Reilly Media, 2020.

\bibitem{voas1995software}
Jeffrey~M. Voas and Keith~W Miller.
\newblock Software testability: The new verification.
\newblock {\em IEEE software}, 12(3):17--28, 1995.

\bibitem{freedman1991testability}
Roy~S Freedman.
\newblock Testability of software components.
\newblock {\em IEEE transactions on Software Engineering}, 17(6):553--564,
  1991.

\bibitem{binder1994design}
Robert~V Binder.
\newblock Design for testability in object-oriented systems.
\newblock {\em Communications of the ACM}, 37(9):87--101, 1994.

\bibitem{gao2005component}
Jerry Gao and M-C Shih.
\newblock A component testability model for verification and measurement.
\newblock In {\em 29th Annual International Computer Software and Applications
  Conference (COMPSAC'05)}, volume~2, pages 211--218. IEEE, 2005.

\bibitem{al2010quality}
Rafa~E Al-Qutaish.
\newblock Quality models in software engineering literature: an analytical and
  comparative study.
\newblock {\em Journal of American Science}, 6(3):166--175, 2010.

\bibitem{mouchawrab2005measurement}
Samar Mouchawrab, Lionel~C Briand, and Yvan Labiche.
\newblock A measurement framework for object-oriented software testability.
\newblock {\em Information and software technology}, 47(15):979--997, 2005.

\bibitem{485220}
A.~Bertolino and L.~Strigini.
\newblock On the use of testability measures for dependability assessment.
\newblock {\em IEEE Transactions on Software Engineering}, 22(2):97--108, 1996.

\bibitem{10.5555/861448}
Jerry~Zayu Gao, Jacob Tsao, Ye~Wu, and Taso H.-S. Jacob.
\newblock {\em Testing and Quality Assurance for Component-Based Software}.
\newblock Artech House, Inc., USA, 2003.

\bibitem{sej.1990.0011}
Richard Bache.
\newblock Measures of testability as a basis for quality assurance.
\newblock {\em Software Engineering Journal}, 5:86--92(6), March 1990.

\bibitem{10.5555/2161638}
Glenford~J. Myers, Corey Sandler, and Tom Badgett.
\newblock {\em The Art of Software Testing}.
\newblock Wiley Publishing, 3rd edition, 2011.

\bibitem{4145039}
B.~Cornelissen, A.~van Deursen, L.~Moonen, and A.~Zaidman.
\newblock Visualizing testsuites to aid in software understanding.
\newblock In {\em 2007 11th European Conference on Software Maintenance and
  Reengineering}, pages 213--222, Los Alamitos, CA, USA, mar 2007. IEEE
  Computer Society.

\bibitem{10.1007/3-540-47993-7_10}
Yoonsik Cheon and Gary~T. Leavens.
\newblock A simple and practical approach to unit testing: The jml and junit
  way.
\newblock In Boris Magnusson, editor, {\em ECOOP 2002 --- Object-Oriented
  Programming}, pages 231--255, Berlin, Heidelberg, 2002. Springer Berlin
  Heidelberg.

\bibitem{1251026}
P.~Runeson and A.~Andrews.
\newblock Detection or isolation of defects? an experimental comparison of unit
  testing and code inspection.
\newblock In {\em 14th International Symposium on Software Reliability
  Engineering, 2003. ISSRE 2003.}, pages 3--13, 2003.

\bibitem{Deursen01refactoringtest}
Arie~Van Deursen, Leon Moonen, Alex Bergh, and Gerard Kok.
\newblock Refactoring test code.
\newblock In {\em Proceedings of the 2nd International Conference on Extreme
  Programming and Flexible Processes in Software Engineering (XP2001}, pages
  92--95, 2001.

\bibitem{159342}
Ieee standard glossary of software engineering terminology.
\newblock {\em IEEE Std 610.12-1990}, pages 1--84, 1990.

\bibitem{ISO9126}
ISO/IEC.
\newblock {\em ISO/IEC 9126. Software engineering -- Product quality}.
\newblock ISO/IEC, 2001.

\bibitem{10.5555/800253.807736}
B.~W. Boehm, J.~R. Brown, and M.~Lipow.
\newblock Quantitative evaluation of software quality.
\newblock In {\em Proceedings of the 2nd International Conference on Software
  Engineering}, ICSE '76, page 592–605, Washington, DC, USA, 1976. IEEE
  Computer Society Press.

\bibitem{bache1990measures}
Richard Bache and Monika Mullerburg.
\newblock Measures of testability as a basis for quality assurance.
\newblock {\em Software Engineering Journal}, 5(2):86--92, 1990.

\bibitem{46876}
S.G. Stockman, A.R. Todd, and G.A. Robinson.
\newblock A framework for software quality measurement.
\newblock {\em IEEE Journal on Selected Areas in Communications},
  8(2):224--233, 1990.

\bibitem{e9292cadbbd244e1993beacd21c53aa1}
John McManus and Trevor Wood-Harper.
\newblock Software engineering: A quality management perspective.
\newblock {\em The TQM Magazine}, 19(4):315--327, 2007.

\bibitem{544352}
V.R. Basili, L.C. Briand, and W.L. Melo.
\newblock A validation of object-oriented design metrics as quality indicators.
\newblock {\em IEEE Transactions on Software Engineering}, 22(10):751--761,
  1996.

\bibitem{10.5555/1875255}
Alain Abran.
\newblock {\em Software Metrics and Software Metrology}.
\newblock Wiley-IEEE Computer Society Pr, 2010.

\bibitem{10.5555/559784}
Stephen~H. Kan.
\newblock {\em Metrics and Models in Software Quality Engineering}.
\newblock Addison-Wesley Longman Publishing Co., Inc., USA, 2nd edition, 2002.

\bibitem{10.5555/2851535}
Ian Sommerville.
\newblock {\em Software Engineering}.
\newblock Pearson, 10th edition, 2015.

\bibitem{bruntink2004predicting}
Magiel Bruntink and Arie Van~Deursen.
\newblock Predicting class testability using object-oriented metrics.
\newblock In {\em Source Code Analysis and Manipulation, Fourth IEEE
  International Workshop on}, pages 136--145. IEEE, 2004.

\bibitem{badri2012empirical}
Mourad Badri and Fadel Toure.
\newblock Empirical analysis of object-oriented design metrics for predicting
  unit testing effort of classes.
\newblock {\em Journal of Software Engineering and Applications},
  5(7):513--526, 2012.

\bibitem{baudry2002testability}
Benoit Baudry, Yves Le~Traon, and Gerson Suny{\'e}.
\newblock Testability analysis of a uml class diagram.
\newblock In {\em Proceedings Eighth IEEE Symposium on Software Metrics}, pages
  54--63. IEEE, 2002.

\bibitem{jungmayr2002identifying}
Stefan Jungmayr.
\newblock Identifying test-critical dependencies.
\newblock In {\em International Conference on Software Maintenance, 2002.
  Proceedings.}, pages 404--413. IEEE, 2002.

\bibitem{briand2003investigating}
Lionel~C Briand, Yvan Labiche, and Hong Sun.
\newblock Investigating the use of analysis contracts to improve the
  testability of object-oriented code.
\newblock {\em Software: Practice and Experience}, 33(7):637--672, 2003.

\bibitem{khatri2013improving}
Sujata Khatri, Rajender~Singh Chhillar, and VB~Singh.
\newblock Improving the testability of object-oriented software during testing
  and debugging processes.
\newblock {\em arXiv preprint arXiv:1308.3320}, 2013.

\bibitem{terragni2020measuring}
Valerio Terragni, Pasquale Salza, and Mauro Pezz{\`e}.
\newblock Measuring software testability modulo test quality.
\newblock In {\em Proceedings of the 28th International Conference on Program
  Comprehension}, pages 241--251, 2020.

\bibitem{shirabad2000supporting}
Jelber~Sayyad Shirabad, Timothy~C Lethbridge, and Stan Matwin.
\newblock Supporting maintenance of legacy software with data mining
  techniques.
\newblock In {\em Proceedings of the 2000 conference of the Centre for Advanced
  Studies on Collaborative research}, page~11, 2000.

\bibitem{tian2015characteristics}
Yuan Tian, Meiyappan Nagappan, David Lo, and Ahmed~E Hassan.
\newblock What are the characteristics of high-rated apps? a case study on free
  android applications.
\newblock In {\em 2015 IEEE international conference on software maintenance
  and evolution (ICSME)}, pages 301--310. IEEE, 2015.

\bibitem{sun2017revisiting}
Chen Sun, Abhinav Shrivastava, Saurabh Singh, and Abhinav Gupta.
\newblock Revisiting unreasonable effectiveness of data in deep learning era.
\newblock In {\em Proceedings of the IEEE international conference on computer
  vision}, pages 843--852, 2017.

\bibitem{coles2016pit}
Henry Coles, Thomas Laurent, Christopher Henard, Mike Papadakis, and Anthony
  Ventresque.
\newblock Pit: a practical mutation testing tool for java.
\newblock In {\em Proceedings of the 25th international symposium on software
  testing and analysis}, pages 449--452, 2016.

\bibitem{jia2010analysis}
Yue Jia and Mark Harman.
\newblock An analysis and survey of the development of mutation testing.
\newblock {\em IEEE transactions on software engineering}, 37(5):649--678,
  2010.

\bibitem{andrews2005mutation}
James~H Andrews, Lionel~C Briand, and Yvan Labiche.
\newblock Is mutation an appropriate tool for testing experiments?
\newblock In {\em Proceedings of the 27th international conference on Software
  engineering}, pages 402--411, 2005.

\bibitem{just2014mutants}
Ren{\'e} Just, Darioush Jalali, Laura Inozemtseva, Michael~D Ernst, Reid
  Holmes, and Gordon Fraser.
\newblock Are mutants a valid substitute for real faults in software testing?
\newblock In {\em Proceedings of the 22nd ACM SIGSOFT International Symposium
  on Foundations of Software Engineering}, pages 654--665, 2014.

\bibitem{nasrabadi2021learning}
Morteza~Zakeri Nasrabadi and Saeed Parsa.
\newblock Learning to predict software testability.
\newblock In {\em 2021 26th International Computer Conference, Computer Society
  of Iran (CSICC)}, pages 1--5. IEEE, 2021.

\bibitem{malhotra2021threshold}
Ruchika Malhotra and Anjali Sharma.
\newblock Threshold benchmarking for feature ranking techniques.
\newblock {\em Bulletin of Electrical Engineering and Informatics},
  10(2):1063--1070, 2021.

\end{thebibliography}

\end{document}